\documentstyle[12pt]{article}

\newcommand{\be}{\begin{equation}}
\newcommand{\ee}{\end{equation}}
\newcommand{\dlt}{\delta}

\newcommand{\ra}{\rightarrow}

\newcommand{\bt}{\beta}

\newcommand{\prt}{\partial}

\begin{document}

\begin{center}
{\Large{\bf Extrapolation of power series by self-similar factor and root
approximants} \\ [5mm]

V.I. Yukalov$^{1,2}$ and S. Gluzman$^3$} \\ [5mm]

{\it
$^1$ Freie Universit\"at Berlin, Fachbereich Physik,
Institut f\"ur Theoretische Physik, \\
WE 2, Arnimallee 14, D-14195 Berlin, Deutschland \\ [3mm]

$^2$Bogolubov Laboratory of Theoretical Physics, \\
Joint Institute for Nuclear Research, Dubna 141980, Russia\\ [3mm]

$^3$Generation 5 Mathematical Technologies Inc., Corporate Headquarters, \\
243 Consumers Road, Suite 800, Toronto, Ontario M2J 4WB, Canada}

\end{center}

\vskip 1.5cm

\begin{abstract}

The problem of extrapolating the series in powers of small variables
to the region of large variables is addressed. Such a problem is typical
of quantum theory and statistical physics. A method of extrapolation is
developed based on self-similar factor and root approximants, suggested
earlier by the authors. It is shown that these approximants and their
combinations can effectively extrapolate power series to the region of
large variables, even up to infinity. Several examples from quantum and
statistical mechanics are analysed, illustrating the approach.

\end{abstract}

\vskip 1cm


\vskip 1cm


\newpage

\section{Introduction}

A common problem in physics is how to extrapolate the series, derived
by means of perturbation theory in powers of a small variable, to large
values of this variable [1]. The known methods of extrapolation are
mainly numerical and rather complicated [1,2]. Recently, being based
on the self-similar approximation theory [3--8] and employing the algebraic 
self-similar renormalization and self-similar bootstrap [9--11], we have 
derived new types of approximants allowing for the summation of power 
series, which are the self-similar exponential approximants [11--14], 
self-similar root approximants [15--17], and self-similar factor approximants 
[18,19], which, for brevity, could be called superexponentials, superroots, 
and superfactors, respectively.

For the purpose of extrapolation, superexponentials are useful for
describing the processes with exponential characteristics, such as
financial time series [20--23] or the development of ruptures [24].
Extrapolation with the help of superexponentials was employed for
analyzing and predicting financial markets [25--29] as well as ruptures
and fractures of materials [30--33].

The processes, characterized by power-law behaviour, require for their
extrapolation either the superroots or superfactors. Actually, both these 
constructions are the realization of the same resummation procedure under 
slightly different initial assumptions. It is the aim of the present paper 
to consider the extrapolation of asymptotic series by means of superroots 
and superfactors, and also by constructing their combinations. We especially 
concentrate on the problem of extrapolation of series, valid for small values 
of variable, to the region, where this variable tends to infinity. We 
illustrate the consideration by several examples typical of quantum and 
statistical physics.

\section{Factor and Root Approximants}

Suppose we are looking for a function $f(x)$ of a real variable $x$, which
is a solution of a complicated physical problem. Let this problem be so much
complicated that it can be treated only by means of perturbation theory
resulting in an expansion
\be
\label{1}
f(x) \simeq \sum_n a_n\; x^n \qquad (x\ra 0)
\ee
in powers of the asymptotically small variable $x$. But assume that our final
goal is to find the behaviour of function $f(x)$ at very large $x$, such that
$x\ra\infty$. This is exactly the extreme variant of the extrapolation problem:
how to extrapolate the finite series
\be
\label{2}
f_k(x) = \sum_{n=0}^k a_n\; x^n \; ,
\ee
having sense solely for $x\ra 0$, to the region, where $x\ra\infty$?

An effective summation of the power series (2) can be done in the frame
of the self-similar approximation theory [3--8], resulting in self-similar
root [15--17] or factor [18,19] approximants. We shall not repeat here
the derivation of these approximants, since all details of the derivation
procedure can be found in our earlier papers, but we shall just present the
resulting formulas that will be employed in the following sections.

If $2k$ terms of a perturbative series are available, its effective summation
can be done by the factor approximant
\be
\label{3}
f^*_{2k}(x) = a_0 \; \prod_{i=1}^k (1 + A_i\; x)^{n_i} \; ,
\ee
where $2k$ control parameters $A_i$ and $n_i$ can be obtained by the
accuracy-through-order procedure, expanding Eq. (3) in powers of $x$ up
to order $2k$ and comparing the like-order coefficients with those of the
perturbative expansion $f_{2k}(x)$. This is the way of defining the even-order
factor approximants. The odd-order approximants can be constructed in several
ways [19], the simplest of which is by leaving untouched the zero-order term,
which yields
\be
\label{4}
f^*_{2k+1}(x) = a_0 + a_1 \; x \; \prod_{i=1}^k (1 + A_i\; x)^{n_i} \; .
\ee
Here and in what follows, the labelling of the factor approximants is done
according to the nontrivial number of perturbative terms required for the
approximant construction. Thus, an $f_k^*(x)$ approximant requires the
knowledge of terms up to order $k$.

Instead of prescribing one of the terms  in the small-$x$ expansion (1), one
can impose a restriction on the behaviour of the sought function at large
$x\ra\infty$, provided this behaviour is known, where
\be
\label{5}
f(x) \sim x^\bt \qquad (x\ra\infty) \; .
\ee
This restriction for the even approximant (3) yields the condition
$$
\sum_{i=1}^k n_i = \bt \; ,
$$
while for the odd approximant (4), one has
$$
1 + \sum_{i=1}^k n_i = \bt \; .
$$
Generally, several such conditions could be invoked, if one would know several
terms describing the behaviour of $f(x)$ at large $x\ra\infty$.

A $k$-order root approximant can be defined through $k$ terms of the large-$x$
expansion [15--17]. Here we shall consider another way of defining root
approximants, by expanding them in powers of $x$ and comparing such expansions
with the given perturbative series (1). Then an even-order root approximant
\be
\label{6}
f^{(j)}_{2k}(x) = a_0\left ( \ldots \left  (  (1+A_1\; x)^{n_1}+
A_2\; x^2\right )^{n_2} + \ldots + A_k\; x^k \right )^{n_k}
\ee
requires the knowledge of $2k$ nontrivial terms of the small-$x$ series (1).
The upper index in approximant (6) labels different possible solutions for the
control parameters $A_i$ and $n_i$, since the accuracy-through-order procedure
for Eq. (6) is not unique. This is contrary to the way of defining the superroot
control parameters from the large-$x$ expansions, which is a uniquely defined
procedure [34]. An odd-order root approximant, with the control parameters
defined from the small-$x$ expansion, reads as
\be
\label{7}
f^{(j)}_{2k+1} = a_0 + a_1\; x \left (\ldots \left( ( 1+ A_1\; x)^{n_1} +
A_2\; x^2\right )^{n_2} + \ldots + A_k\; x^k\right )^{n_k} \; .
\ee

Note that the control parameters $A_i$ and $n_i$ in each of the approximants
given by Eqs. (3), (4), (6), and (7) are of course different, and we use the
same letters for their notation in order to avoid too cumbersome nomenclature.
In the present paper, we shall also consider the hybrid approximants combining
the forms of factor and root approximants.

In those cases, when the subsequent self-similar approximants $f_k^*(x)$ display
substantial oscillations, it proved effective to introduce the averaged form
\be
\label{8}
\overline{f}_k^*(x) = \sum_{i=1}^k p_i(x)\; f_i^*(x) \; ,
\ee
where $f_i^*(x)$ are weighted with the probabilities
\be
\label{9}
p_i(x) \equiv \frac{|m_i(x)|^{-1}}{\sum_{j=1}^N |m_j(x)|^{-1} }\; ,
\ee
in which $N$ is the number of available approximants and
\be
\label{10}
m_i(x) \equiv \frac{\dlt f_i^*(x)}{\dlt f_0^*(x)} =
\frac{\prt f_i^*(x)}{\prt x}\; {\Huge /} \; \frac{\prt f_0^*(x)}{\prt x}
\ee
are the mapping multipliers [35], where we may set $f_0^*(x)\equiv a_0+a_1x$.
Averages (8) can be defined for both factor as well as root approximants.

\section{Hybrid Factor-Root Approximants}

Here we show the way of constructing factor approximants, root approximants,
and their various combinations. We illustrate this on the case of an 
asymptotic expansion obtained from the simple function
\be
\label{11}
f(x) = \frac{1}{x}\; \ln(1+x) \; .
\ee
Expanding this function in powers of $x$ gives series (1) with the 
coefficients 
\be
\label{12}
a_n = \frac{(-1)^n}{n+1} \; .
\ee
Retaining in this expansion the terms up to 6-th order, allows us to define
the factor approximant
\be
\label{13}
f_6^*(x) = (1+A_1\; x)^{n_1}(1+A_2\; x)^{n_2}(1+A_3\; x)^{n_3} \; ,
\ee
whose control parameters are to be found from the accuracy-through-order
procedure, which yields
$$
A_1=0.9767\; , \qquad A_2=0.6261\; , \qquad A_3=0.1830\; ,
$$
$$
n_1=-0.3503\; , \qquad n_2=-0.1935\; , \qquad n_3=-0.2009\; .
$$

At large $x$, approximant (13) behaves as
\be
\label{14}
f(x) \simeq A_1^{n_1} A_2^{n_2} A_3^{n_3}\; x^{n_1+n_2+n_3} \qquad
(x\ra\infty) \; ,
\ee
where
$$
A_1^{n_1} A_2^{n_2} A_3^{n_3} = 1.5528 \; , \qquad n_1+n_2+n_3=-0.7447\; .
$$

The root approximant of the same order is
\be
\label{15}
f_6^{(j)}(x) =  \left\{ \left [ ( 1 + A_1\; x)^{n_1} + A_2\; x^2
\right ]^{n_2} + A_3\; x^3 \right \}^{n_3} \; .
\ee
Expanding approximant (15) in powers of $x$, comparing this expansion with
series (1), and equating the coefficients at like powers, we get the system
of equations possessing $28$ solutions. However, a natural restriction, 
limiting the number of admissible solutions, is the requirement that 
approximant (15) be real. Then only 5 solutions remain. The upper index 
$j$ in Eq. (15) enumerates these solutions, which we shall analyse in turn.

For the solution $f_6^{(1)}(x)$ we have
$$
A_1=0.9059\; , \qquad A_2=0.0728\; , \qquad A_3=0.0932\; ,
$$
$$
n_1=2.2152\; , \qquad n_2=0.9424\; , \qquad n_3=-0.2644\; .
$$
At large $x$, we find
\be
\label{16}
f_6^{(1)}(x) \simeq A_3^{n_3}\; x^{3n_3} \qquad (x\ra\infty) \; ,
\ee
with
$$
A_3^{n_3} = 1.8728 \; , \qquad 3n_3= -0.7931\; .
$$

For another solution $f_6^{(2)}(x)$, the parameters are
$$
A_1=2.0426\; , \qquad A_2=1.2353\; , \qquad A_3=0.1938\; ,
$$
$$
n_1=1.0002\; , \qquad n_2=1.0031\; , \qquad n_3=-0.2440\; .
$$
The asymptotic behaviour at large $x$ is the same as in Eq. (16) but with
$$
A_3^{n_3}=1.4924 \; , \qquad 3n_3 = -0.7320 \; .
$$

The next solution $f_6^{(3)}(x)$ has the parameters
$$
A_1=1.2509\; , \qquad A_2=0.2772\; , \qquad A_3=0.0135\; ,
$$
$$
n_1=1.0616\; , \qquad n_2=1.7956\; , \qquad n_3=-0.2097\; .
$$
The asymptotic behaviour at large $x$ is
\be
\label{17}
f_6^{(3)}(x) \simeq A_2^{n_2n_3}\; x^{2n_2n_3} \qquad (x\ra\infty) \; ,
\ee
where
$$
A_2^{n_2n_3} = 1.6211 \; , \qquad 2n_2 n_3 = -0.7531 \; .
$$

The parameters for $f_6^{(4)}(x)$ are
$$
A_1=1.1673\; , \qquad A_2=0.1682\; , \qquad A_3=-0.0545\; ,
$$
$$
n_1=0.8629\; , \qquad n_2=1.5029\; , \qquad n_3=-0.3303\; .
$$
The asymptotic form at large $x$ is the same as in Eq. (17), but with
$$
A_2^{n_2n_3} = 2.4227 \; , \qquad 2n_2 n_3 = -0.9928 \; .
$$

For the approximant $f_6^{(5)}(x)$, we have
$$
A_1=1.0761\; , \qquad A_2=0.2497\; , \qquad A_3=0.2389\; ,
$$
$$
n_1=1.9117\; , \qquad n_2=1.0210\; , \qquad n_3=-0.2381\; .
$$
The behaviour at large $x$ is of type (16), but with
$$
A_3^{n_3}=1.4062 \; , \qquad 3n_3 = -0.7143 \; .
$$
All approximants $f_6^{(j)}(x)$ are close to each other.

The hybrid approximants of the same order can be defined as follows. 
A possible form is
\be
\label{18}
f_6^{(j)}(x) = \left [ (1 + A_1\; x)^{n_1} + A_2\; x^2 \right ]^{n_2}
(1+A_3\; x)^{n_3} \; .
\ee
Defining the parameters $A_i$ and $n_i$ by the accuracy-through-order
procedure, we again meet the problem of multiple solutions. And again we
reject those solutions that result in complex-valued or divergent functions
(18), since the sought solutions must be finite and real for all $0\leq 
x<\infty$. The remaining approximants are again close to each other. Thus, 
for $f_6^{(6)}(x)$ we have
$$
A_1=0.8053\; , \qquad A_2=0.1131\; , \qquad A_3=0.9767\; ,
$$
$$
n_1=0.9966\; , \qquad n_2=-0.1970\; , \qquad n_3=-0.3501\; .
$$
The large-$x$ behaviour is
\be
\label{19}
f_6^{(6)}(x) \simeq A_2^{n_2} A_3^{n_3}\; x^{2n_2+n_3} \qquad
(x\ra\infty) \; ,
\ee
where
$$
A_2^{n_2} A_3^{n_3} = 1.5490 \; , \qquad 2n_2 + n_3 = -0.7441 \; .
$$

One more solution gives $f_6^{(7)}(x)$ of type (18), but with the parameters
$$
A_1=1.2532\; , \qquad A_2=0.2303\; , \qquad A_3=0.7320\; ,
$$
$$
n_1=1.0251\; , \qquad n_2=-0.2425\; , \qquad n_3=-0.2574\; .
$$
At large $x$, one has the same behaviour as in Eq. (19), but with
$$
A_2^{n_2}A_3^{n_3} = 1.5471\; , \qquad 2n_2+n_3=-0.7424 \; .
$$

Another hybrid approximant can be written as
\be
\label{20}
f_6^{(j)}(x) = \left [  ( 1 + A_1\; x)^{n_1}\; (1+A_2\; x)^{n_2} +
A_3\; x^3 \right ]^{n_3} \; .
\ee
Defining the parameters by means of the accuracy-through-order procedure,
we require that approximant (20) be real  and finite. Typical parameters are
$$
A_1=0.8724\; , \qquad A_2=0.3360\; , \qquad A_3=0.0197\; ,
$$
$$
n_1=1.7396\; , \qquad n_2=0.3551 \; , \qquad n_3=-0.3054\; .
$$
For large $x$, this yields the same form as in Eq. (16), but with
$$
A_3^{n_3}=3.3180\; , \qquad 3n_3=-0.9162\; .
$$

Approximants $f_6^*(x)$ and $f_6^{(j)}(x)$ are of comparable accuracy.
The existence of multiple solutions for the control parameters is somewhat
compensated by the mutual closeness of the approximants corresponding to
different parametric solutions. The accuracy can be essentially improved
by invoking the minimal difference condition [9].

\section{Quartic Anharmonic Oscillator}

Let us consider the one-dimensional quartic anharmonic oscillator with the
Hamiltonian
\be
\label{21}
H = -\; \frac{1}{2}\; \frac{d^2}{dx^2} + \frac{1}{2}\; x^2 + g\; x^4 \; ,
\ee
in which $x\in(-\infty,+\infty)$ and $g\geq 0$. The related ground-state
energy, obtained by means of perturbation theory, is the asymptotic series
\be
\label{22}
E(g) \simeq \sum_n a_n\; g^n \qquad (g\ra 0) \;
\ee
in powers of the coupling $g$. The coefficients $a_n$ can be found in
Refs. [19,36]. The strong-coupling asymptotic behaviour is
\be
\label{23}
E(g) \simeq 0.667986\; g^{1/3} \qquad (g\ra\infty) \; .
\ee
The convergence of the factor approximants $E^*_{2k}(g)$ was shown in [19].
Here we shall compare the factor, root, and hybrid approximants with each
other.

We shall consider expansion (22) up to the $6$-th order in $g$. The
corresponding factor approximant is
\be
\label{24}
E_6^*(g) = \frac{1}{2}\; (1 + A_1\; g)^{n_1}\; (1+A_2\; g)^{n_2}\;
(1+ A_3\; g)^{n_3} \; .
\ee
The control parameters are uniquely defined from the accuracy-through-order
procedure, which yields
$$
A_1=26.4702\; , \qquad A_2=12.4688\; , \qquad A_3=3.8380\; ,
$$
$$
n_1=1.8017 \times 10^{-3}\; , \qquad n_2=0.0547\; , \qquad n_3=0.2005\; .
$$
The strong-coupling limit of Eq. (24) is
\be
\label{25}
E_6^*(g) \simeq \frac{1}{2}\; A_1^{n_1} A_2^{n_2} A_3^{n_3} \;
g^{n_1+n_2+n_3} \qquad (g\ra\infty) \; ,
\ee
where
$$
\frac{1}{2}\;  A_1^{n_1} A_2^{n_2} A_3^{n_3} =0.7561\; , \qquad
n_1+n_2+n_3 =0.2570 \; .
$$
Comparing Eqs. (25) and (23), we see that the amplitude in Eq. (25) is
predicted with an error of $13\%$, and the power, with an error $-23\%$.

Dealing with the root and hybrid approximants, with the control parameters
defined by the accuracy-through-order procedure, we, as always, confront
the problem of nonuniqueness of solutions. Of course, we shall again reject
those solutions which do not guarantee that the related approximant be
real-valued and finite for finite $g$. Moreover, we shall present below only
those root and hybrid approximants of the given order, whose strong-coupling
limit is closer to the $g^{1/3}$ law.

Note, first, that the best 5-th order root approximant
\be
\label{26}
E_5^{(j)}(g) = \frac{1}{2} + \frac{3}{4}\left [ (1+A_1\; g)^{n_1} +
A_2\; g^2 \right ]^{n_2} \; ,
\ee
with the parameters
$$
A_1=24.1009\; , \qquad A_2=125.3648\; , \qquad
n_1=0.8859\; , \qquad n_2=-0.1639\; ,
$$
has the strong-coupling limit
\be
\label{27}
E_5^{(j)} \simeq \frac{3}{4}\; A_2^{n_2}\; g^{2n_2+1} \qquad
(g\ra\infty) \; ,
\ee
where
$$
\frac{3}{4}\; A_2^{n_2} =0.3397 \; , \qquad 2n_2+1 =0.6721 \; .
$$
Hence, the amplitude is given with an error of $-49\%$ and the power, with
an error of $102\%$.

The 6-th order root approximant is
\be
\label{28}
E_6^{(j)}(g) = \frac{1}{2}\left\{ \left [ (1+A_1\; g)^{n_1} + A_2\; g^2
\right ]^{n_2} + A_3\; g^3 \right \}^{n_3} \; .
\ee
The best accuracy is provided by the parameters
$$
A_1=16.0451\; , \qquad A_2=52.5504\; , \qquad A_3=37.0388\; ,
$$
$$
n_1=0.8682 \; , \qquad n_2=5.4769\; , \qquad n_3=0.0197\; .
$$
In the strong-coupling limit, this gives
\be
\label{29}
E_6^{(1)} \simeq \frac{1}{2}\; A_2^{n_2n_3} \; g^{2n_2n_3} \qquad
(g\ra\infty) \; ,
\ee
with
$$
\frac{1}{2}\; A_2^{n_2n_3} = 0.7660\; , \qquad 2n_2n_3=0.2154 \; .
$$
The error of the amplitude is $15\%$ and that of the power is $-35\%$.

Another solution for the parameters, corresponding to form (28), is
$$
A_1=26.6927\; , \qquad A_2=234.0099\; , \qquad A_3=695.5007\; ,
$$
$$
n_1=0.9638 \; , \qquad n_2=0.8934\; , \qquad n_3=0.0653\; .
$$
This results in the strong-coupling behaviour as
\be
\label{30}
E_6^{(2)}(g) \simeq \frac{1}{2}\; A_3^{n_3}\; g^{3n_3} \qquad
(g\ra\infty) \; ,
\ee
with
$$
\frac{1}{2}\; A_3^{n_3}=0.7664\; , \qquad 3n_3=0.1958 \; .
$$
The errors of the amplitude and power are $15\%$ and $-41\%$, respectively.

The best hybrid approximant of the form
\be
\label{31}
E_6^{(3)}(g) = \frac{1}{2}\; \left [ (1+A_1\; g)^{n_1} + A_2\; g^2
\right ]^{n_2} \; (1+A_3\; g)^{n_3}
\ee
possesses the parameters
$$
A_1=30.9204\; , \qquad A_2=245.4475\; , \qquad A_3=4.1366\; ,
$$
$$
n_1=0.9754 \; , \qquad n_2=0.0207\; , \qquad n_3=0.2120\; .
$$
The related strong-coupling limit is
\be
\label{32}
E_6^{(3)}(g) \simeq \frac{1}{2}\; A_2^{n_2} A_3^{n_3}\; g^{2n_2+n_3}
\qquad (g\ra\infty) \; ,
\ee
where
$$
\frac{1}{2}\; A_2^{n_2} A_3^{n_3} = 0.7570\; , \qquad 2n_2+n_3=0.2533.
$$
The amplitude and power errors are $13\%$ and $-24\%$, respectively.

Another hybrid approximant of the type
\be
\label{33}
E_6^{(4)}(g) = \frac{1}{2}\; \left [ (1+A_1\; g)^{n_1}\;
(1+A_2\; g)^{n2} + A_3\; g^3 \right ]^{n_3} \; ,
\ee
with the best parameters
$$
A_1=7.7952\; , \qquad A_2=25.9485\; , \qquad A_3=97.8519\; ,
$$
$$
n_1=2.1670 \; , \qquad n_2=0.0263\; , \qquad n_3=0.0853\; ,
$$
has the same strong-coupling limit as (30), but with
$$
\frac{1}{2}\; A_3^{n_3}=0.7394\; , \qquad 3n_3=0.2560 \; .
$$
The extrapolation of the amplitude is done with an error of $11\%$ and that
of the power, with an error of $-23\%$.

In this way, the extrapolation accuracy of the best hybrid approximant (33)
is practically the same as that of the factor approximant (24). But the latter
has the advantage of being uniquely defined. The accuracy of factor approximants
can be further improved by considering either the simple Cesaro averages for
the neighbouring approximants, as
\be
\label{34}
\frac{1}{2}\; \left [ E_{k-1}^*(g) + E_k^*(g) \right ] \; ,
\ee
or the weighted averages of type (8). As we have checked, the accuracy of the
weighted averages
\be
\label{35}
\overline E_k^*(g) =  p_{k-1}(g) E_{k-1}^*(g) + p_k(g) E_k^*(g) \; ,
\ee
with probabilities (9), is essentially better than that of the simple 
averages (34). One more possibility for improving the accuracy is to use the 
minimal-difference condition for the subsequent approximants [9,37].

\section{Boxed Quantum Particle}

The problem of defining the ground state energy of a quantum particle in a
one-dimensional box can be formulated [38,39] as the problem of finding the
limit of the function
\be
\label{36}
E(g) = \frac{\pi^2}{128} \; \left ( \frac{1}{2} + \frac{16}{\pi^4g^2} +
\frac{1}{2}\; \sqrt{1+ \frac{64}{\pi^4g^2} } \right )
\ee
as $g\ra\infty$. The energy is written here in dimensionless units. One has
\be
\label{37}
E(\infty) = \frac{\pi^2}{128} = 0.077106 \; .
\ee
We shall analyse how this value can be extrapolated from the weak-coupling
expansion
\be
\label{38}
E(g) \simeq \frac{1}{8\pi^2 g^2} \; \sum_n a_n\; g^n \qquad (g\ra 0) \; .
\ee
The initial coefficients of the latter expansion are
$$
a_0=1 \; , \quad a_1 =\frac{\pi^2}{4}=2.467401 \; , \quad
a_2=\frac{\pi^4}{32}=3.044034 \; , \quad 
a_3=\frac{\pi^6}{512}=1.877713 \; ,
$$
$$
a_4=0 \; , \qquad a_5=-0.714478\; , \qquad a_6=0\; , 
\qquad a_7=0.543724 \; ,
$$
$$
a_8=0\; , \qquad a_9=-0.517223\; , \qquad a_{10}=0 \; , \qquad 
a_{11}=0.551056 \; , \qquad a_{12}=0\; .
$$

The even factor approximants are
\be
\label{39}
E_{2k}^*(g) = \frac{1}{8\pi^2 g^2} \; \prod_{i=1}^k ( 1 + A_i\; g)^{n_i} \; .
\ee
To guarantee a finite limit, as $g\ra\infty$, we impose the restriction
\be
\label{40}
\sum_{i=1}^k n_i = 2 \; .
\ee
The control parameters $A_i$ and $n_i$ are uniquely defined from the
accuracy-through-order procedure.

In the fourth order, we have $E_4^*(g)$ with the parameters
$$
A_1=0.30843+0.81602\; i \; , \qquad A_2=A_1^* \; , \qquad
n_1=1-1.13389\; i \; , \qquad n_2=n_1^* \; .
$$
Then we get
\be
\label{41}
E_4^*(\infty) = 0.14968\; ,
\ee
with an error of $90\%$.

For the factor approximant $E_6^*(g)$, we find
$$
A_1=0.44119\; , \qquad A_2=0.08783+1.02776\; i \; , \qquad
A_3=A_2^* \; ,
$$
$$
n_1=1.43469 \; , \qquad n_2=0.28266-0.86829\; i \; , \qquad n_3=n_2^* \; .
$$
As a result,
\be
\label{42}
E_6^*(\infty) = 0.05257 \; ,
\ee
whose error is $-32\%$.

The approximant $E_8^*(g)$ possesses the parameters
$$
A_1=0.27455+0.39441\; i \; , \qquad A_2=A_1^* \; , \qquad
A_3=0.03388+1.11925\; i \; , \qquad A_4=A_3^* \; ,
$$
$$
n_1=0.88578-0.77500\; i \; , \qquad n_2=n_1^* \; , \qquad
n_3=0.11422-0.60842\; i \; , \qquad n_4=n_3^* \; .
$$
This yields
\be
\label{43}
E_8^*(\infty) = 0.10285 \; ,
\ee
with an error of $33\%$.

For the approximant $E_{10}^*(g)$, the parameters are
$$
A_1=0.14438+0.65311\; i \; , \qquad A_2=A_1^* \; ,
$$
$$
A_3=0.01626+1.16211\; i \; , \qquad A_4=A_3^* \; , \qquad
A_5=0.29557 \; ,
$$
$$
n_1=0.30513-0.70804\; i \; , \qquad n_2=n_1^* \; ,
$$
$$
n_3=0.06119-0.46375\; i \; , \qquad n_4=n_3^*\; , \qquad
n_5=1.26735\; .
$$
The limiting value is
\be
\label{44}
E_{10}^*(\infty) = 0.06201 \; ,
\ee
which deviates from the exact value (37) by an error of $-20\%$. Thus, the
sequence of the approximants $E_{2k}^*(\infty)$ displays numerical convergence
to the exact result (37).

Constructing the root approximants, with the control parameters defined by
the accuracy-through-order procedure, we select, as early, the best approximants.
And again, we impose the condition that the energy at $g\ra\infty$ is finite. For
example, the root approximant
\be
\label{45}
E_4^{(1)}(g) = \frac{1}{8\pi^2g^2}\; \left [ (1+A_1\; g)^{n_1} + A_2\; g^2
\right ]^{n_2}
\ee
is assumed to satisfy the conditions
\be
\label{46}
n_1<2\; , \qquad n_2=1
\ee
guaranteeing that $E_4^{(1)}(\infty)$ is finite. The parameters of Eq. (45) are
$$
A_1=3.4826\; , \qquad A_2=4.2965\; , \qquad n_1=0.7085\; , \qquad n_2=1 \; .
$$
The sought limit is
\be
\label{47}
E_4^{(1)}(\infty) =0.0544 \; ,
\ee
which has an error of $-30\%$.

As an odd root approximant, we consider
\be
\label{48}
E_5^{(1)}(g) = \frac{1}{8\pi^2g^2}\; \left\{ 1 +
\frac{\pi^2}{4}\; g\; \left [ (1+A_1\; g)^{n_1} + A_2\; g^2 \right ]^{n_2}
\right \} \; ,
\ee
under the restriction
\be
\label{49}
n_1<2 \; , \qquad n_2 = \frac{1}{2} \; .
\ee
The best solution for the parameters results in
\be
\label{50}
E_5^{(1)}(\infty)=0.0648 \; ,
\ee
which has an error of $-16\%$.

The accuracy of the root approximants is about the same as that of the factor
approximants. Though the best root approximants are slightly better than the
factor approximants of the same order, this advantage is spoiled by the problem
of multiplicity of solutions for the root parameters, while the parameters for
the factor approximants are uniquely defined.

Note that the method of Pad\'e approximants also incorporates the problem of
nonuniquely defined approximants, since for a given order $k$ of series (2),
one may construct a table of $k$ Pad\'e-approximants $f_{[M/N]}$, with all $M$
and $N$ satisfying the equality $M+N=k$. One often inclines to use the diagonal
Pad\'e approximants [2], which are not necessarily the most accurate. Thus, for
series (38) the best Pad\'e approximant of $6$-th order is $E_{[4/2]}(g)$,
yielding
$$
E_{[4/2]}(\infty) = 0.03855 \; ,
$$
whose error is $-50\%$.

The accuracy of the factor approximants can be drastically improved by
constructing their averages. For instance, even for the simple Cesaro averages
we have
$$
\frac{1}{2}\; \left ( E_4^* + E_6^* \right ) = 0.10113 \qquad (31\%) \; ,
$$
$$
\frac{1}{2}\; \left ( E_6^* + E_8^* \right ) = 0.07771 \qquad (0.8\%) \; ,
$$
$$
\frac{1}{2}\; \left ( E_8^* + E_{10}^* \right ) = 0.08243 \qquad (7\%) \; ,
$$
where $E_k^*\equiv E_k^*(\infty)$ and the percentage errors are shown in
brackets. Even better is the accuracy of the weighted averages (8) involving
two nearest neighbours. We find
$$
\overline E_6^*=0.0778 \qquad (0.9\%) \; ,
$$
$$
\overline E_8^*=0.0774 \qquad (0.4\%) \; ,
$$
$$
\overline E_{10}^*=0.0776 \qquad (0.6\%) \; ,
$$
In this way, it looks that the most convenient technique would be by
constructing the factor approximants and forming their weighted averages.

\section{Fluctuating Fluid Membrane}

A problem, mathematically very similar to that considered in the previous
section, is the determination of the pressure of a tensionless membrane
between walls [40]. This pressure, in dimensionless units, can be presented
as the strong-coupling limit
\be
\label{51}
P(\infty) = \lim_{g\ra\infty} P(g)
\ee
of a function $P(g)$ that can be calculated for small $g\ra 0$ as a series
\be
\label{52}
P(g) \simeq \frac{1}{4g^2} \; \sum_n a_n\; g^n \qquad (g\ra 0)\; .
\ee
Monte Carlo estimates give [41,42] the  value
\be
\label{53}
P(\infty) = 0.0798\pm 0.0003 \; .
\ee
The coefficients in series (52) can be written [39] as
$$
a_0=0.0506606\; , \qquad a_1=0.125000\; , \qquad a_2=0.154213 \; ,
$$
$$
a_3=0.105998\; , \qquad a_4=0.026569\; , \qquad a_5=-0.034229\; ,
\qquad a_6=-0.083251 \; .
$$

The even factor approximants are defined by the formula
\be
\label{54}
P^*_{2k}(g) = \frac{a_0}{4g^2}\; \prod_{i=1}^k (1+A_i\; g)^{n_i} \; ,
\ee
in which, in order to guarantee the finite limit (51), one should set
\be
\label{55}
\sum_{i=1}^k n_i = 2 \; .
\ee
Then one has
\be
\label{56}
P_{2k}^*(\infty) = \frac{a_0}{4}\; \prod_{i=1}^k A_i^{n_i} \; .
\ee

However, the factor approximants (54), under condition (55), do not provide
good accuracy for this problem. For example, $P_2^*(\infty)=0.019$, which is
too small. To the contrary, $P_4^*(\infty)=0.312$ is too large.

The root approximants here work much better. Thus the approximant
\be
\label{57}
P_4^{(1)}(g) = \frac{a_0}{4g^2} \left [ (1+A_1\; g)^{n_1} + A_2\; g^2
\right ]^{n_2} \; ,
\ee
with the parameters
$$
A_1=3.5607\; , \qquad A_2=4.3928\; , \qquad n_1=0.6930\; , \qquad n_2=1\; ,
$$
gives $P_4^{(1)}(\infty)=0.0556$, where an error is $-30\%$.

The odd root approximant
\be
\label{58}
P_5^{(1)}(g) = \frac{a_0}{4g^2}\; \left\{ 1 + A g \left [
(1+A_1\; g)^{n_1} + A_2\; g^2 \right ]^{n_2} \right \} \; ,
\ee
where
$$
A=2.4674\; , \qquad A_1=3.7056\; , \qquad A_2=4.7456 \; ,
$$
$$
n_1=0.6659\; , \qquad n_2=0.5 \; ,
$$
yields $P_5^{(1)}(\infty)=0.0681$, with an error of $-15\%$.

The 6-th order root approximant
\be
\label{59}
P_6^{(1)}(g) = \frac{a_0}{4g^2} \; \left ( \left\{ \left [
(1+A_1\; g)^{n_1} + A_2\; g^2 \right ]^{n_2} + A_3\; g^3 \right\}^{n_3}
\right ) \; ,
\ee
with
$$
A_1=4.8198\; , \qquad A_2=11.4910\; , \qquad A_3=13.2536 \; ,
$$
$$
n_1=0.9893\; , \qquad n_2=0.7762\; , \qquad n_3=2/3 \; ,
$$
results in $P_6^{(1)}(\infty)=0.0709$, which has an error of $-11\%$.

The odd root approximant
\be
\label{60}
P_7^{(1)}(g) = \frac{a_0}{4g^2}\; \left ( 1 + Ag \left \{ \left [
(1+A_1\; g)^{n_1} + A_2\; g^2 \right ]^{n_2} + A_3\; g^3 \right \}^{n_3}
\right ) \; ,
\ee
in which
$$
A=2.4674\; , \qquad A_1=4.8298 \; , \qquad A_2=11.9969 \; ,
\qquad A_3=15.0003 \; ,
$$
$$
n_1=0.9994\; , \qquad n_2=0.7668 \; , \qquad n_3=1/3 \; ,
$$
gives $P_7^{(1)}(\infty)=0.07707$, which is accurate to within $-3.4\%$.

Recall that the root approximants are not uniquely defined from
the accuracy-through-order procedure. Here only the best of them are
presented. Different solutions for the control parameters usually lead
to the approximants that are close to each other. For instance, another
solution for form (60) would give $P_7^{(2)}(g)$ with the parameters
$$
A=2.4674\; , \qquad A_1=2.3970\; , \qquad A_2=6.1342\; ,
\qquad A_3=14.9918 \; ,
$$
$$
n_1=2.0166\; , \qquad n_2=0.7657 \; , \qquad n_3=1/3 \; .
$$
This gives practically the same limit $P_7^{(2)}(\infty)=0.07706$ as
$P_7^{(1)}(\infty)$.

Pad\'e approximants, invoked for this problem, either contain divergencies,
or yield principally incorrect results with the negative values of pressure.

\section{One-Dimensional Antiferromagnet}

The ground-state energy of the one-dimensional Heisenberg antiferromagnet
is known exactly [43],
\be
\label{61}
E=-0.4431 \; .
\ee
From another side, this energy can be considered as the limit
$$
E=\lim_{t\ra\infty} E(t)
$$
of the temporal energy $E(t)$. The latter can be calculated for small $t$,
yielding the so-called $t$-expansion [44]
\be
\label{62}
E(t) \simeq -\; \frac{1}{4}\; \sum_n a_n\; t^n \qquad (t\ra 0) \; ,
\ee
in which
$$
a_0=1\; , \qquad a_1=4\; , \qquad a_2=-8\; , \qquad
a_3=-\; \frac{16}{3}\; , \qquad a_4=64\; .
$$

The even 4-order factor approximant extrapolating expansion (62) is
\be
\label{63}
E_4^*(t) = -\; \frac{1}{4}\; (1+A_1\; t)^{n_1}\; (1+A_2\; t)^{n_2} \; ,
\ee
which in the long-time limit gives
$$
E_4^*(t) \simeq -\; \frac{1}{4}\; A_1^{n_1} A_2^{n_2}\; t^{n_1+n_2}
\qquad (t\ra\infty) \; .
$$
The parameters of Eq. (63) are uniquely defined by the
accuracy-through-order procedure, with the condition
$$
n_1+n_2=0 \; ,
$$
guaranteeing the finiteness of $E_4^*(\infty)$. This gives
$E_4^*(\infty)=-0.570$, with an error of $29\%$, as compared to the
Hulthen result [43].

The odd factor approximant
\be
\label{64}
E_5^*(t) = -\; \frac{1}{4}\; \left [ 1 +4t (1+A_1\; t)^{n_1}
(1+A_2\; t)^{n_2} \right ]
\ee
has the limiting behaviour
$$
E_5^*(t) \simeq -\; \frac{1}{4}\; \left ( 1 + 4A_1^{n_1}A_2^{n_2}\;
t^{n_1+n_2+1}\right ) \qquad (t\ra \infty) \; ,
$$
with the restriction
$$
n_1+n_2+1=0 \; .
$$
From here, $E_5^*(\infty)=-0.4452$, which is accurate to within $0.5\%$.

The best root approximant
\be
\label{65}
E_5^{(j)}(t) = -\; \frac{1}{4}\; \left \{ 1 + 4t\left [ (1+A_1\; t)^{n_1}
+A_2\; t^2\right ]^{n_2}\right \} \; ,
\ee
with the asymptotic form
$$
E_5^{(j)}(t) \approx -\; \frac{1}{4} \left ( 1 + 4A_2^{n_2}\;
t^{2n_2+1}\right ) \qquad (t\ra\infty) \; ,
$$
requires the condition
$$
2n_2+1=0 \; .
$$
The related limit is $E_5^{(j)}(\infty)=-0.4743$, whose error is $7\%$.

The diagonal Pad\'e approximant $E_{[2/2]}(t)$ gives $E_{[2/2]}(\infty)
=-0.3289$, with an error of $26\%$.

\section{Combined Factor-Exponential Approximants}

When the asymptotic behaviour of a function at large $x\ra\infty$ is
a combination of the power-law and exponential dependence, the extrapolation
of the corresponding series at small $x\ra 0$ can be done by combining the
factor and exponential approximants. We shall illustrate this by considering
the Airy function satisfying the Airy equation
\be
\label{66}
\frac{d^2}{dx^2}\; {\rm Ai}(x) - x{\rm Ai}(x) = 0 \; .
\ee
The solution to this equation at small $x$ can be written as a series
\be
\label{67}
{\rm Ai}(x) \simeq \sum_n a_n \; x^n \qquad (x\ra 0) \; ,
\ee
substituting which in Eq. (66) gives the coefficients
$$
a_0=0.355028\; , \qquad a_1=-0.258819\; , \qquad a_2=0 \; ,
\qquad a_3=\frac{a_0}{6} \; ,
$$
$$
a_4=\frac{a_1}{12}\; , \qquad a_5=0\; , \qquad a_6=\frac{a_0}{180} \; ,
\qquad a_7=\frac{a_1}{504} \; , \qquad a_8=0\; , \; \quad \ldots
$$
At large $x$, the Airy function behaves as
\be
\label{68}
{\rm Ai}(x) \simeq \frac{a_0}{2\sqrt{\pi}}\; x^{-1/4}\; \exp\left ( -\;
\frac{2}{3}\; x^{3/2} \right ) \qquad (x\ra\infty) \; .
\ee
The combined factor-exponential approximant of fifth order can be represented
as
\be
\label{69}
A_5^*(x) = a_0 (1+A_1\; x)^{n_1}\; \exp\left\{ bx^2(1+A_2\; x)^{n_2}
\right\} \; ,
\ee
where the methods of Refs. [11,12] are involved.
The parameters, found from the accuracy-through-order procedure, are
$$
A_1=1.480028\; , \;\; A_2=1.400808\; , \;\; b=-0.805208\; , \;\;
n_1=-0.492565\; , \;\; n_2=-0.505184\; .
$$
The large-$x$ behaviour of approximant (69) is
\be
\label{70}
A_5^*(x)\simeq a_0 A_1^{n_1}\; x^{n_1} \;
\exp\left ( bA_2^{n_2}\; x^{n_2+2} \right ) \; ,
\ee
where
$$
bA_2^{n_2}=-0.697141\; , \qquad n_2+2=1.494816\; ,
$$
which are very close to the values $-2/3$ and $3/2$, respectively.
Approximant (69) represents the exact Airy function very well. Thus,
the error for $x<10$ is less than $1\%$.

\section{Discussion of Other Problems}

The considered technique can be applied to any problem requiring an
extrapolation of small-variable asymptotic series to the large-variable
region. We have accomplished such an extrapolation for a variety of problems. 
However, we would not like to overload this paper by the description of 
calculational details corresponding to the related problems. Instead, we 
shall just briefly summarize the cases we have analysed.

The luminescent intensity of donor-acceptor recombination [45] was
extrapolated by means of the factor approximants in Ref. [19]. Now,
we have also considered the extrapolation with the help of the root
approximants. The best variants of the latter are not better than
the factor approximants of the same order. The accuracy can be
improved by invoking averages (8). The best Pad\'e approximants
are much less accurate than the self-similar approximants.

The nonlinear Schr\"odinger equation is met in several physical
applications. A very important and interesting application is the
description of coherent fields of trapped Bose atoms, when the
corresponding nonlinear equation is called the Gross-Pitaevskii
equation [46--50]. Self-similar approximants  for the ground-state
wave function and energy of the latter equation were considered in
Refs. [16,51] and for the whole spectrum of excited energies in Refs.
[52,53]. Here, we have compared the extrapolation procedure, based
on both factor and root approximants, for the ground-state energy.
We find that for this problem the factor approximants are more
accurate than the root approximants.

The factor approximants are also convenient for describing critical
phenomena. Thermodynamic characteristics in the vicinity of the
critical point exhibit the behaviour typical of one of the factors
[54,55]. Earlier, we checked the applicability of subsequent factor
approximants to different critical phenomena [18,19]. Now we have shown
that the hybrid factor-root approximants can also be used for describing
critical phenomena. We have checked this for the elliptic integral with
logarithmic singularity [19] and for the so-called (2+1) dimensional
Ising model [56]. The accuracy of the approximants can be improved by
involving the minimal-difference condition [9,37].

Critical behaviour may also appear in the solutions to nonlinear
differential equations [23,57--59]. We studied this effect for the
Ruina-Dietrich equation by using the factor approximants [18]. As we
have now checked this critical behaviour can also be described by the
hybrid factor-root approximants.

In the majority of the cases we have investigated, Pad\'e approximants
are essentially less accurate than the self-similar approximants. Often,
Pad\'e approximants are not applicable at all, being divergent or
qualitatively incorrect. In particular cases, it is possible to fit
the correct behaviour of a sought function by making manipulations
with Pad\'e approximants raising them to some fractional powers [60].
However, such a way of fitting is too arbitrary to serve as a serious
method. In addition, this fitting actually results in constructions
analogous to self-similar root approximants, though slightly spoiled
and not so symmetric.

\section{Conclusion}

A power series in powers of an asymptotically small variable $x\ra 0$
can be effectively extrapolated to the region, where this variable is
large, and even to the limiting case $x\ra\infty$. This can be done by
means of the self-similar factor and root approximants. Hybrid approximants,
combining factors and roots, can also be used for extrapolation. The control
parameters can be defined from the accuracy-through-order relations. This
procedure yields unique solutions for the factor approximants but multiple
solutions for root approximants. Fortunately, the multiplicity of solutions
for the root parameters is often not as dangerous, leading to approximants
that are very close to each other.

The extrapolation by the factor approximants is preferable, being
more accurate, when the sought function either increases to infinity,
as $x\ra\infty$, or diminishes to zero. When the function tends, as
$x\ra\infty$, to a nonzero finite value, the accuracy of the root
approximants can become higher than that of the factor approximants.

The usage of the complete root-factor technique could be justified in 
those cases when the number of available terms in a small-variable asymptotic
expansion is limited, if the derivation of the higher-order terms is too 
costly or even impossible at all. The trouble with multiple solutions 
could be overcome by imposing additional restrictions on the behaviour of 
solutions at infinity.

A very important feature of the root and hybrid root-factor approximants 
is their nontrivial behaviour at infinity allowing for defining the 
so-called corrections to scaling, while the direct usage of the factor 
approximants yields only the leading exponent. This problem will be 
studied in detail in a separate paper.

\vskip 5mm

{\bf Acknowledgement}

\vskip 3mm

We appreciate helpful discussions with B. Kastening, to whom we also are 
thankful for providing us the coefficient $a_6$ for the expansion (52) of
Section 6.

One of the authors (V.I.Y.) is very much grateful to E.P. Yukalova for many 
useful discussions and criticism. He also appreciates financial support from 
the German Research Foundation (DFG grant Be 142/72-1).

\end{document}